\documentstyle[12pt]{article}
\newcommand{\ie}{{\em i.e.}}

\newcommand{\lsim}{\stackrel{\scriptstyle <}{{ }_\sim}}
\newcommand{\bb}{\overline}
\newcommand{\rta}{\rightarrow}
\newcommand{\be}{\begin{equation}}
\newcommand{\ee}{\end{equation}}
\newcommand{\ben}{$$}
\newcommand{\een}{$$}
\newcommand{\h}[1]{\mbox{$h_{#1}$}}
\newcommand{\de}[1]{\mbox{$\delta_{#1}$}}
\newcommand{\dep}[1]{\mbox{$\delta'_{#1}$}}
\newcommand{\gevu}{GeV^{-1}}
\newcommand{\gevd}{GeV^{-2}}
\newcommand{\cl}{\quad\quad (90\%\;\mbox{C.L.})}
\newcommand{\qp}{\; .}
\newcommand{\qc}{\; ,}
\newcommand{\hc}{\mbox{h.c.}}
\begin{document}
\thispagestyle{empty}
\begin{flushright}
{\parbox{3.5cm}{
UAB-FT-378

November, 1995

hep-ph/9511297
}}
\end{flushright}
\vspace{3cm}
\begin{center}
\begin{large}
\begin{bf}
HIGGS TRIPLET EFFECTS IN PURELY LEPTONIC PROCESSES \\
\end{bf}
\end{large}
\vspace{1cm}
J.A. Coarasa, A. M\'endez and J. Sol\`a \\
\vspace{0.25cm}
Grup de F\'{\i}sica Te\`orica and IFAE \\
\vspace{0.25cm}
Universitat Aut\`onoma de Barcelona\\
08193 Bellaterra (Barcelona), Catalonia, Spain\\
\end{center}
\vspace{0.3cm}
\begin{center}
{\bf ABSTRACT}
\end{center}
\begin{quotation}
We consider the effect of complex Higgs triplets on purely leptonic
processes and survey the experimental constraints on the mass and
couplings of their single and double charge members. Present day
experiments tolerate values of the Yukawa couplings of these scalars
at the level of the standard electroweak gauge couplings. We show that
the proposed measurement of the ratio $R_{LCD}=\sigma (\nu_{\mu}e)/
[\sigma (\bb\nu_{\mu}e) + \sigma (\nu_e e )]$ would allow to explore
a large region of the parameter space inaccessible to the usual ratio
$R=\sigma (\nu_{\mu}e)/\sigma (\bb\nu_{\mu}e)$.

\end{quotation}

\newpage

Although the Standard Model (SM) of the strong and electroweak
interactions seems to agree very well with the experimental data,
the Higgs sector still remains practically unexplored and one can
therefore consider different extensions.
Here we will focus our attention on Higgs triplets
with non zero hypercharge. They are a common feature in models where
the left-handed neutrinos acquire a Majorana mass through the Higgs
mechanism. This is the case, for instance, of the classical version
of the Left--Right Symmetric extension of the SM \cite{lrsm}.

Higgs triplets under the standard $SU(2)_L$ gauge group have some
very specific properties. Their vacuum expectation values have to be
small in order not to spoil the agreement between the theoretical and
the experimental values of the electroweak $\rho$
parameter \cite{refgrimen}. Moreover,
since they have two units of weak hypercharge, apart from the
electrically neutral scalar ($h^0$) there are also singly charged
($h^+$) and
doubly charged ($h^{++}$) ones. This has an important
phenomenological consequence, namely, that they {\em can not couple to
quarks\/} and therefore their dominant effects are to be observed only
in purely leptonic processes. Moreover, these effects are not
necessarily small
because the Higgs triplets do not participate in the generation of the
lepton {\em Dirac\/} masses and therefore their Yukawa couplings can
be large.

Another feature displayed by scalar triplets (and, more generally, by
any unmixed higher dimension scalar multiplet) is that they exhibit
equal mass-squared spacing ordered by the scalar field electric
charges \cite{refweiler,refgunion}. This implies that, in general,
$m_{h^{++}}>m_{h^{+}}$.

Several experimental constraints on the couplings and masses of the
$h^{++}$ have been considered in the
literature \cite{refgunion,refbarger,refswartz,refmohap}.
They include the constraints arising
from the value of the anomalous magnetic moment of the electron and
the muon, the angular distribution in the
Bhabha scattering and the
experimental limits on muonium--antimuonium transitions
and non standard muon decays.
The effects of the singly charged member of the
triplet, $h^+$, have not been considered so extensively. The
existence of such field would affect the purely leptonic neutrino
processes and precise measurements in this sector, like those of the
CHARM II collaboration \cite{refcharm2},
impose additional constraints. In this respect,
there has been a proposal by the Large Cerenkov Detector (LCD) Project
collaboration \cite{reflcd} of an experiment at LAMPF that would
allow the measurement of the ratio
\be
R_{LCD} \equiv \frac{\sigma
(\nu_{\mu} e )}{\sigma (\bb\nu_{\mu} e ) + \sigma (\nu_e e )} \qc
\label{eqlcd}
\ee
with a precision of 2\%\ (corresponding to a 0.9\%\ precision in the
measurement of $\sin^2\theta_W$). Such high precision is due to the
particular setup of this experiment which will be almost free of
systematic uncertainties \cite{reflcd}, making it competitive with
the standard CHARM II ratio
$R=\sigma (\nu_{\mu}e)/\sigma (\bb\nu_{\mu}e)$ from which the value of
$\sin^2\theta_W$ is obtained with an accuracy at the level of 3.6\% .
Therefore, the ratio $R_{LCD}$ could be a suitable tool to probe
physics beyond the SM. This is the case, for instance, of
supersymmetric effects \cite{refcms}.
The existence of $h^\pm$ would also affect this ratio and more
stringent bounds on its couplings and mass might be imposed from the
measured value.

In this paper  we shall determine the region in the parameter space
where the effects of the $h^\pm$ bosons could be observed in the
measurement of $R_{LCD}$. This requires first an updating of the
experimental constraints on the $h^\pm$ mass and Yukawa couplings.
At the end we shall comment on the implications for the $h^{\pm\pm}$
mass and couplings.

The term of the lagrangian describing the interaction of the triplet
with the leptons can be written as
\be
{\cal L} = i \sum_{i,j=e,\mu ,\tau} g_{ij} \left( {{\Psi_i}^T_L }\;
{\cal C} \tau_2
\Delta {{\Psi_j}}_L \right) + \hc \qc
\label{eqlagrangian}
\ee
where ${\cal C}$ is the charge conjugation matrix, ${\Psi_i}_L$ are the
standard lepton doublets and $\Delta$ is the
scalar triplet written in the usual matrix form,
\ben
\Delta=\left( {\matrix{h^+/\sqrt{2} & h^{++}\cr
                       h^0 & -h^+/\sqrt{2} \cr}} \right) \qp
\een

The interactions described by the lagrangian of eq.~(\ref{eqlagrangian})
do not conserve the lepton family numbers in general. They do conserve,
however, the {\em total} lepton number, $L$, if the value $L=-2$ is
assigned to the scalar triplet $\Delta$.

We assume the coupling constants $g_{ij}$ to be real.
The relevant Feynman rules are shown in fig.~1 in terms of
$\h{ij}\equiv (g_{ij}+g_{ji})/2$. (Notice that, by definition,
$\h{ij}=\h{ji}$.)

Defining $\de{ij}\equiv \h{ij}/m_{h^+}$,
we list now the constraints on $\de{ij}$ arising from different
experimental measurements.

\vspace{0.25cm}
\noindent
{\bf a.  Restrictions from the decay $\mu\rta e \gamma$}

Parametrizing the $\mu e \gamma$ vertex in the usual way,
\ben
\Lambda^{\lambda} + \Lambda^{\lambda}_A \gamma_5 \qc
\een
with
\begin{eqnarray}
\Lambda^{\lambda}=F_1\gamma^{\lambda}+
\frac{F_2}{m_\mu+m_e} i \sigma^{\lambda \nu} q_{\nu}+
\frac{F_3}{m_\mu+m_e} q^{\lambda}\nonumber\qc \\
\Lambda^{\lambda}_A=G_1\gamma^{\lambda}+
\frac{G_2}{m_\mu+m_e} i \sigma^{\lambda \nu} q_{\nu}+
\frac{G_3}{m_\mu+m_e} q^{\lambda}\qc\nonumber
\end{eqnarray}
the decay width, $\Gamma\left( \mu\rta e
\gamma\right)$, in the limit $m_e << m_\mu$ can be written as
\ben
\Gamma\left( \mu\rta e
\gamma\right)=\frac{\alpha m_{\mu}}{2}
\left(\left|F_2\right|^2+\left|{G_2}\right|^2\right) \qc
\een
and the corresponding branching ratio is
\be
B(\mu\rta e \gamma)=\frac{96 \pi^2 \alpha}{G_F^2 m^4_\mu}
\left(\left|F_2\right|^2+\left|{G_2}\right|^2\right) \qp
\label{eqbr}
\ee
The $h^-$ contribution to $F_2$, $G_2$ arising from the diagram of
fig.~2.a is
\be
F_2=G_2
=\frac{m^2_\mu}{192 \pi^2}\left[
\de{e\mu}\left(\de{e e}+\de{\mu\mu}\right)+\de{e\tau}\de{\mu\tau}
\right] \qp
\label{eqf2g2}
\ee
{}From the experimental bound \cite{pdg94}
\ben
B(\mu\rta e \gamma)
<4.9\times 10^{-11}\cl\qc
\een
using eqs.(\ref{eqbr}) and (\ref{eqf2g2}) we obtain
\ben
\left|\de{e\mu}\left(\de{e e}+\de{\mu\mu}\right)+
\de{e\tau}\de{\mu\tau}\right|<
4.2\times 10^{-8} \gevd\cl\qp
\een

\vspace{0.25cm}
\noindent
{\bf b. Restrictions from the decay $\mu\rta e \nu_e \bb\nu_\mu$
and neutrino oscillation experiments}

The existence of $h^-$ would allow the non standard decay
$\mu\rta e \nu_e \bb\nu_\mu$ through the diagram of
fig.~2.b. The corresponding width is
\ben
\Gamma\left( \mu\rta e \nu_e \bb\nu_\mu\right)=
\frac{m^5_\mu}{1536 \pi^3}\de{ee}^2\de{\mu\mu}^2 \qp
\een
{}From the experimental limit
\cite{pdg94}
\ben
\frac{\Gamma\left( \mu\rta e \nu_e \bb\nu_\mu\right)}{
\Gamma\left( \mu\rta e \bb\nu_e\nu_\mu \right)}<0.012\cl\qc
\een
the following bound can be obtained
\ben
\left|\de{ee}\de{\mu\mu}\right|<3.6\times 10^{-6} \gevd\cl\qp
\een
(If we further assume $e-\mu$ universality then we would have
$\left|\de{ee}\right|=\left|\de{\mu\mu}\right|<1.9\times 10^{-3}$.)

Recently, it has been pointed out in ref.~\cite{nuosc} that the
results of the KARMEN experiment \cite{karmen} on neutrino
oscillations lead to the bound
\be
\left(\frac{G_N}{G_F}\right)^2 < 3.1\times 10^{-3} \gevd\cl\qc
\label{eqgn}
\ee
where $G_F$ is the Fermi constant and
$G_N$ is the coupling constant of an effective four fermion
interaction of the form
\ben
{\cal L} = G_N \left(\bb\mu\gamma_\lambda P_L e\right)
\left(\bb\nu_\mu \gamma^\lambda P_L \nu_e \right) + \hc \qc
\een
with $P_L=(1-\gamma_5)/2$.
The amplitude of the diagram of fig.~2.b can be
written this form after a suitable Fierz reordering with
$G_N\equiv \de{ee}\de{\mu\mu}$. Then, from eq.~(\ref{eqgn})
we can obtain the more restrictive bound
\be
\left|\de{ee}\de{\mu\mu}\right|<6.5\times 10^{-7} \gevd\cl\qp
\label{cotaosc}
\ee
(Again, assuming $e-\mu$ universality we would have
$\left|\de{ee}\right|=\left|\de{\mu\mu}\right|<8.1\times 10^{-4}$.)

\vspace{0.25cm}
\noindent
{\bf c. Restrictions from anomalous magnetic moments}

By explicit calculation of the diagrams of fig.~2.c we find the
following contribution to the anomalous magnetic moment,
$ a_i\equiv (g-2)/2$, of the charged lepton $i$,
\be
\Delta a_i=
-\frac{ m^2_i}{48 \pi} \sum_{j=e,\mu,\tau} \de{ij}^2 \qp
\label{gm2p}
\ee
The theoretical values of $a_e$ and $a_\mu$ are \cite{kinoshita}
\begin{eqnarray}
a_e^{th}   &=& (1159652460\pm 127\pm 75)\times 10^{-12}\qc\nonumber \\
a_\mu^{th} &=& (11659202\pm 20)\times 10^{-10} \qc\nonumber
\end{eqnarray}
and the corresponding experimental values are \cite{pdg94}
\begin{eqnarray}
a_e^{exp}   &=& (1159652193\pm 10)\times 10^{-12}\qc\nonumber \\
a_\mu^{exp} &=& (11659230\pm 84)\times 10^{-10} \qp\nonumber
\end{eqnarray}
{}From these values we
can obtain the 90\%\ C.L. intervals for any non standard contribution
to $a_i$ \footnote{They are given by
$a^{exp}-a^{th} \pm 1.645\, (\sigma^2_{exp}+\sigma^2_{th})^{1/2}$.},
\begin{eqnarray}
-5.1\times 10^{-10}<& \Delta a_e   <& -0.2\times 10^{-10}\qc\nonumber \\
-1.1\times 10^{-8} <& \Delta a_\mu <& 1.7\times 10^{-8}\qp\nonumber
\end{eqnarray}
Since the contribution given by eq.~(\ref{gm2p}) is negative, the
following bounds apply,
\be
\left|\Delta a_e\right| < 5.1\times 10^{-10} \qc \quad
\left|\Delta a_\mu\right| < 1.1\times 10^{-8} \qp
\label{cotadelai}
\ee
Then, the following restrictions are obtained,
\begin{eqnarray}
\sum_{j=e,\mu,\tau} \de{ej}^2 &< 0.29\, \gevd &\cl\qc\nonumber\\
\sum_{j=e,\mu,\tau} \de{\mu j}^2 &< 1.5\times 10^{-4}\gevd &\cl\qc
\nonumber
\end{eqnarray}
from which we can infer the following bounds,
\begin{eqnarray}
\left|\de{ee}\right|,\left|\de{e\tau}\right| &< 0.54\,
\gevu &\cl\qc\nonumber\\
\left|\de{\mu\mu}\right|,\left|\de{\mu e}\right|,\left|\de{\mu\tau}
\right| &< 1.2\times 10^{-2} \gevu &\cl\qp
\label{cotagmd}
\end{eqnarray}

\vspace{0.25cm}
\noindent
{\bf d. Restrictions from the value of $\sin^2\theta_W$
measured in purely leptonic neutrino interactions}

Using the ratio $R=\sigma (\nu_{\mu}e)/\sigma (\bb\nu_{\mu}e)$,
the CHARM II Collaboration \cite{refcharm2} has been able to obtain a
precise value of $\sin^2\theta_W$ from purely leptonic
$\nu_\mu e$ and $\bb\nu_\mu e$ collisions,
\be
x_W \equiv \sin^2\theta_W = 0.2324\pm 0.0083 \qp
\label{eqsw2}
\ee
The SM tree level amplitude for the $\nu_\mu e $ scattering is given by
\be
{\cal M}=i\frac{G_F}{\sqrt 2}\bb e\gamma_\lambda\left[
P_L - 2 x_W \right] e\times
\bb\nu\gamma^\lambda (1-\gamma_5) \nu \qp
\label{eqnme}
\ee
A similar expression with straightforward modifications affecting
only the last
factor of the r.h.s. of eq.~(\ref{eqnme}) can be written for the
$\bb\nu_\mu e$ amplitude.

The two $h^-$ exchanging diagrams of
fig.~2.d add new contributions to the $\nu_\mu e$ and
$\bb\nu_\mu e$ amplitudes respectively.
With a suitable Fierz transformation one can readily see that the net
effect of these $h^-$ contributions consists in replacing the
term inside the square
brackets of eq.~(\ref{eqnme}) by
\begin{eqnarray}
P_L-2 x_W&\rta &
\left( 1+\frac{\de{e\mu}^2}{\sqrt{2}G_F}\right) P_L -
2 x_W\nonumber\\
&\simeq&\left( 1+\frac{\de{e\mu}^2}{\sqrt{2}G_F}\right)
\left[P_L - 2 x_W
\left( 1-\frac{\de{e\mu}^2}{\sqrt{2}G_F}\right)\right] \qp
\label{eqxwrho}
\end{eqnarray}
This replacement induces an apparent shift of $x_W$ and the $\rho$
parameter given by
\ben
\Delta x_W\simeq -\frac{x_W \de{e\mu}^2}{\sqrt{2}G_F}\qc\quad
\Delta \rho= \frac{\de{e\mu}^2}{\sqrt{2}G_F} \qp
\een

The agreement between the results of the CHARM II Collaboration with
those obtained from semi--leptonic $\nu$ interactions (where $h^\pm$
effects would be absent) suggests that $\de{e\mu}^2/\sqrt{2}G_F$ is
small (of the order of a few percent at most) and thus justifies the
use of the approximate form given by the last expression of
eq.~(\ref{eqxwrho}).

The value given by eq.~(\ref{eqsw2}) coincides with the theoretical
prediction, $x_W^{th}=0.2324\pm 0.0012$ (see, e.g., ref.~\cite{holl}).
The 90\%\ C.L. interval for $\Delta x_W$ is therefore
\ben
\left|\Delta x_W\right| < 0.0138 \qc
\een
from which we obtain
\be
\left|\de{e\mu}\right|<9.9\times 10^{-4}\gevu\cl\qp
\label{cotacharm}
\ee

A similar analysis can be carried out in the case of $\nu_e e$
collisions. The experimental study has been done at LAMPF \cite{nueres}
and the obtained experimental value for $x_W$ is
\ben
x_W = 0.249\pm 0.063 \qp
\een
In this case the SM tree level amplitude can be conveniently written as
\ben
{\cal M}=i\frac{G_F}{\sqrt 2}\bb e\gamma_\lambda\left[
-2P_L + (P_L - 2 x_W) \right] e\times
\bb\nu\gamma^\lambda (1-\gamma_5) \nu \qp
\een
The effect of the $h^-$
contribution is again the substitution (\ref{eqxwrho}) with
$\de{e\mu}$ replaced by $\de{e e}$.
The extra term, $-2P_L$, (compared to the $\nu_\mu e$ case) is
due to the $W$ exchange contribution and is unaffected by the $h^-$
exchange. The apparent shift on $x_W$ is now
\ben
\Delta x_W\simeq -\frac{x_W \de{e e}^2}{\sqrt{2}G_F} \qc
\een
and the 90\%\ C.L. interval for $\Delta x_W$ is
\ben
-0.09 < \Delta x_W < 0.12 \qc
\een
which leads to the following bound
\be
\left|\de{e e}\right| < 2.5 \times 10^{-3}\gevu\cl\qp
\label{cotalampf}
\ee

An alternative approach in order to
obtain bounds on non SM physics from the measured values of the
$\sigma (\nu_e e)$ is described in ref. \cite{nueres} . If we write
\ben
\frac{d\sigma (\nu_e e)}{dy} = \sigma_0
\left[A + B (1-y) + C (1-y)^2 \right] \qc
\een
where $\sigma_0 = G^2_F s/4\pi$,
the $h^-$ exchange diagram adds the following  contribution to the
quantity $A$,
\ben
\Delta A = -(2+4x_W) \frac{\de{e e}^2}{\sqrt{2}G_F} \qc
\een
and from the experimental value of $\sigma (\nu_e e)$ the following
bound on $\de{e e}$ can be derived (see ref. \cite{nueres} for
details\footnote{The fact that the $h^-$ exchange diagram modifies the
quantity $A$ is due to the helicity structure in the lagrangian of
eq.~(\ref{eqlagrangian}). A charged Higgs belonging to a {\em doublet}
would, instead, contribute to $C$ (see ref.~\cite{nueres}).}),
\ben
\left|\de{e e}\right| < 7.2 \times 10^{-3}\gevu\cl\qc
\een
which is less restrictive than the bound
of eq.~(\ref{cotalampf}).

\vspace{0.25cm}
The restrictions imposed by the above observables on
$\de{ee}$, $\de{\mu\mu}$ and $\de{e\mu}$ are
summarized in table~\ref{tab1} where only the most restrictive ones,
given by eqs. (\ref{cotaosc}), (\ref{cotagmd}), (\ref{cotacharm}) and
(\ref{cotalampf}), are listed. The combined bounds on $\de{ij}$ are
also shown in fig.~3.

We now turn to the ratio $R_{LCD}$ defined in eq.~(\ref{eqlcd}).
Introducing the quantities
\begin{eqnarray}
\epsilon_-&=&\frac{1}{2}(1-2x_W)\qc\nonumber\\
\epsilon'_-&=&\frac{1}{2}(-1-2x_W)\qc\nonumber\\
\epsilon_+&=&\epsilon'_+=-x_W \qc\nonumber
\end{eqnarray}
the tree level SM value can be written as
\be
R_{LCD}^0 = \frac{\epsilon^2_- + \frac{1}{3}\epsilon^2_+}
    {\epsilon^2_+ + \frac{1}{3}\epsilon^2_- +
    \epsilon'^2_- + \frac{1}{3}\epsilon'^2_+} = 0.142 \qp
\label{eqlcd0}
\ee

This value is modified by the presence of non SM physics
(in our case, the $h^-$ interactions) as well as by
SM radiative corrections. The dominant $h^-$ contributions
to the cross sections appearing in eq.~(\ref{eqlcd}) are given by
the two diagrams of fig.~2.d and the first one
involving the electron neutrino.
The effect of these $h^-$ exchange contributions is taken into
account by simply replacing in eq.~(\ref{eqlcd0}) the quantities
$\epsilon_\pm$ and $\epsilon'_\pm$ by $\tilde\epsilon_\pm$ and
$\tilde\epsilon'_\pm$ respectively, where
\begin{eqnarray}
\tilde\epsilon_+ &= \epsilon_+\; ,\quad
\tilde\epsilon_- &= \epsilon_- + \frac{1}{2}
\frac{\de{e\mu}^2}{\sqrt{2}G_F}\qc\nonumber\\
\tilde\epsilon'_+ &= \epsilon'_+\; ,\quad
\tilde\epsilon'_- &= \epsilon'_- + \frac{1}{2}
\frac{\de{e e}^2}{\sqrt{2}G_F}\qp\nonumber
\end{eqnarray}
Thus, we have
\ben
R_{LCD} = \frac{\tilde\epsilon^2_- + \frac{1}{3}\tilde\epsilon^2_+}
{\tilde\epsilon^2_+ + \frac{1}{3}\tilde\epsilon^2_- +
\tilde\epsilon'^2_- + \frac{1}{3}\tilde\epsilon'^2_+} +
\Delta R_{LCD}^{R.C.}\qc
\een
where $\Delta R_{LCD}^{R.C.}$ is the effect of the SM radiative
corrections which have been computed in ref.~\cite{marciano}. The
result is
\ben
\Delta R_{LCD}^{R.C.} = 0.036\times R_{LCD}^0 = 5.1\times 10^{-3}\qp
\een

In fig.~3.b we plot the contour lines of
$R_{LCD}/R_{LCD}^{SM}$, where
\ben
R_{LCD}^{SM}\equiv R_{LCD}^0 + \Delta R_{LCD}^{R.C.} = 0.147\qc
\een
is the SM prediction (including radiative corrections).

We observe that in a large part of the allowed region in the
$|\de{e e}|$-$|\de{e\mu}|$ plane the deviations of the ratio $R_{LCD}$
from the SM prediction are larger than 2\%\ and therefore detectable.
For some (allowed) values of $\de{e e}$ and $\de{e\mu}$ these
deviations can be even larger than 50\%\ (in the top right corner of
the allowed region of fig.~3.b the value of $R_{LCD}$ would double the
SM one).
On the other hand, if no deviations from the SM were observed the
region above the 1.02 curve would be ruled out.

If the inequality $m_{h^+}<m_{h^{++}}$ holds, as it happens
in practically all Higgs triplet models,
both the measurement of $R_{LCD}$ and the bounds on $\de{ij}$
described above would
also affect the restrictions on the mass and couplings of the doubly
charged scalar, $h^{++}$.
Indeed, defining $\dep{ij}\equiv \h{ij}/m_{h^{++}}$, if the singly
charged scalars are lighter than the doubly charged ones then
$\dep{ij}<\de{ij}$ and therefore any upper bound of $\de{ij}$ would be
also an upper bound of $\dep{ij}$.

The direct bounds on $\dep{ij}$ (without assuming the inequality
$m_{h^+}<m_{h^{++}}$) are obtained from the experimental results
on $(g-2)$, Bhabha scattering, muonium-antimuonium ($M\bb M$)
transitions and the decay $\mu^-\rta e^-e^+e^-$. They
are summarized in the left half of table~\ref{tab2}.

The $(g-2)$ bound is obtained by applying eq.~(\ref{cotadelai}) to the
$h^{++}$ contribution to $a_i$ given by
\cite{refgunion},
\be
\Delta a_i=
-\frac{ m^2_i}{6 \pi} \sum_{j=e,\mu,\tau} \dep{ij}^2 \qp
\label{gm2pp}
\ee
The resulting inequalities are
\begin{eqnarray}
\left|\dep{ee}\right|,\left|\dep{e\tau}\right| &< 0.19\,
\gevu &\cl\qc\nonumber\\
\left|\dep{\mu\mu}\right|,\left|\dep{\mu e}\right|,\left|\dep{\mu\tau}
\right| &< 4.3\times 10^{-3} \gevu &\cl\qp\nonumber
\end{eqnarray}
The remaining bounds are discussed
in ref.~\cite{refswartz}\footnote{The $M\bb M$ bound has been
updated according to the new experimental limit given in
ref.~\cite{pdg94}. The Bhabha scattering and $\mu\rta 3e$ bounds were
also discussed in ref.~\cite{refgunion}. Notice that
our $h_{ee}$, $h_{\mu\mu}$ and
$h_{e\mu}$ are respectively the $g_{ee}$, $g_{\mu\mu}$ and
$g_{e\mu}/2$ of ref.~\cite{refswartz}.}.

If $m_{h^+}<m_{h^{++}}$\ , the bounds of table~\ref{tab1} became also
bounds on $\dep{ij}$. We show them in the right half of
table~\ref{tab2}. In the case
of $\dep{\mu\mu}$, the slight improvement of
the bound is obtained by adding the $h^+$ and $h^{++}$ contributions
to $a_i$, given by eqs. (\ref{gm2p}) and (\ref{gm2pp}) respectively,
and using the inequality $\dep{ij}<\de{ij}$. We have then
\ben
\frac{3m^2_i}{16 \pi} \sum_{j=e,\mu,\tau} \dep{ij}^2 <
\left| \Delta a_i^{(h^+)} + \Delta a_i^{(h^{++})} \right| \qc
\een
with the r.h.s. satisfying the bounds of eq.~(\ref{cotadelai}).

The $\mu\rta 3e$ bound on $|\dep{ee} \dep{e\mu}|$ appearing in the
last line of table~\ref{tab2} is, by far, the most restrictive
one and implies that either $m_{h^{++}}$ is very large or
$h_{ee}h_{e\mu}$ is very small. If the couplings $h_{ij}$ are not
small compared to, say, the $SU(2)_L$ gauge coupling $g$, then
$h^{++}$ must be very massive. In fact, if $h_{ee}h_{e\mu}\sim g^2$
the $h^{++}$ mass would be extremely large ($\sim 100\, TeV$). The
same applies to $m_{h^+}$ and $m_{h^0}$, for if the three masses
were very different there would be unacceptable contributions to the
electroweak $\rho$ parameter \cite{refgunion}.
As a result, the quantities $\de{ij}$ and
$\dep{ij}$ would be very small and Higgs triplet effects
{\em at the tree level} would be hard to detect. Nevertheless, there
could still be sizeable one-loop effects through (oblique)
contributions to the $\rho$ parameter.

If, instead, $h_{ee}h_{e\mu}$ is very small, one can consider two
situations. One possibility is that
both couplings are very small as it happens with the Yukawa
couplings of the SM Higgs to the first two fermion generations, which
are several orders of magnitude smaller than the gauge couplings,
a fact that is considered an unnatural feature of the SM.
Another theoretically more appealing possibility
is that $h_{e\mu}=0$ (\ie , that the Higgs
triplet interactions are diagonal in lepton flavour) and
$h_{ee}\sim h_{\mu\mu}\sim g$. Indeed, one can have this type
of scenario compatible with all the bounds of tables \ref{tab1} and
\ref{tab2}. For instance, if $m_{h^+}\lsim m_{h^{++}}\sim 500\, GeV$,
one could have $h_{ee}h_{\mu\mu}\sim g^2/5$ while fulfilling the $\rho$
parameter constraint. Thus,
extended Higgs sectors as the one considered here are specially
attractive since they can be ``natural'',  entailing a
potentially rich phenomenology beyond the SM.

\newpage
\begin{large}
\noindent
{\bf Acknowledgements}
\end{large}

\vspace{0.3cm}
\noindent
Work supported in part by the CICYT research project AEN95-0882. J.S.
is grateful to W.C. Louis and D.H. White for their hospitality at LAMPF
and for useful discussions on the LCD project.


\vspace{1.5cm}
\begin{center}
\begin{large}
{\bf Figure Captions}
\end{large}
\end{center}
\begin{itemize}
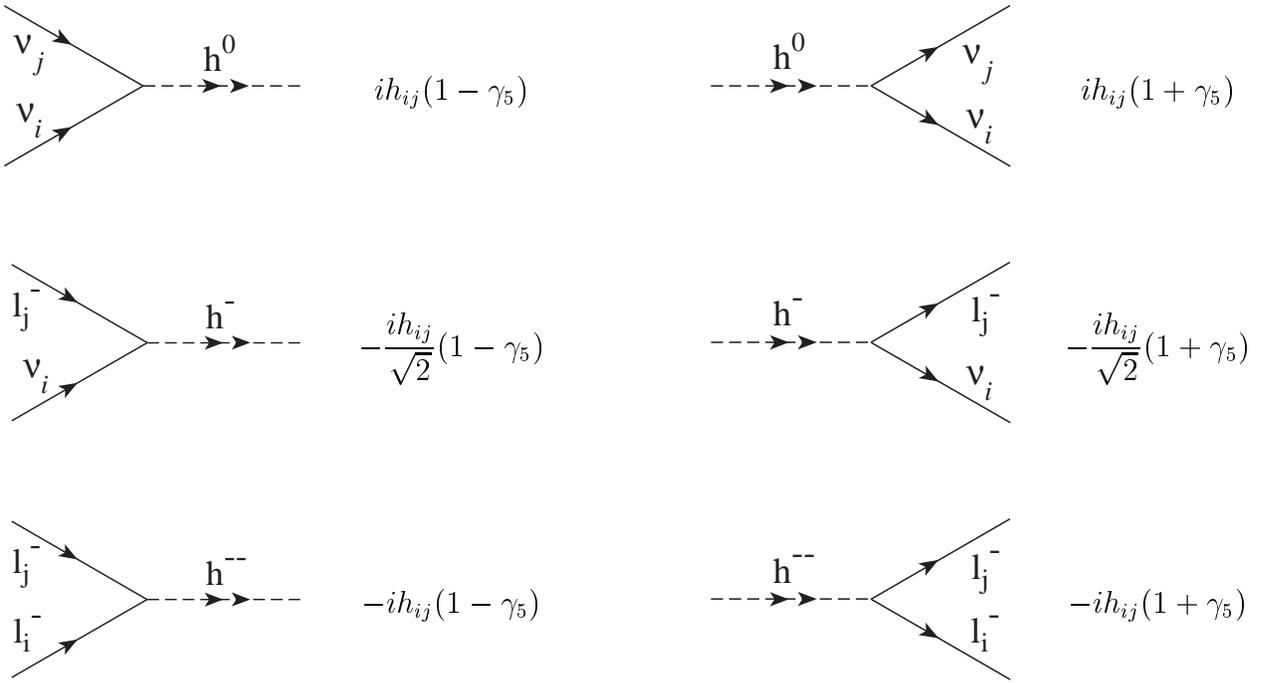
\item[\bf Fig. 1:] Feynman rules corresponding to the interactions
   described by the lagrangian of eq.~(\ref{eqlagrangian}). Each arrow
   indicates one unit of total lepton number.
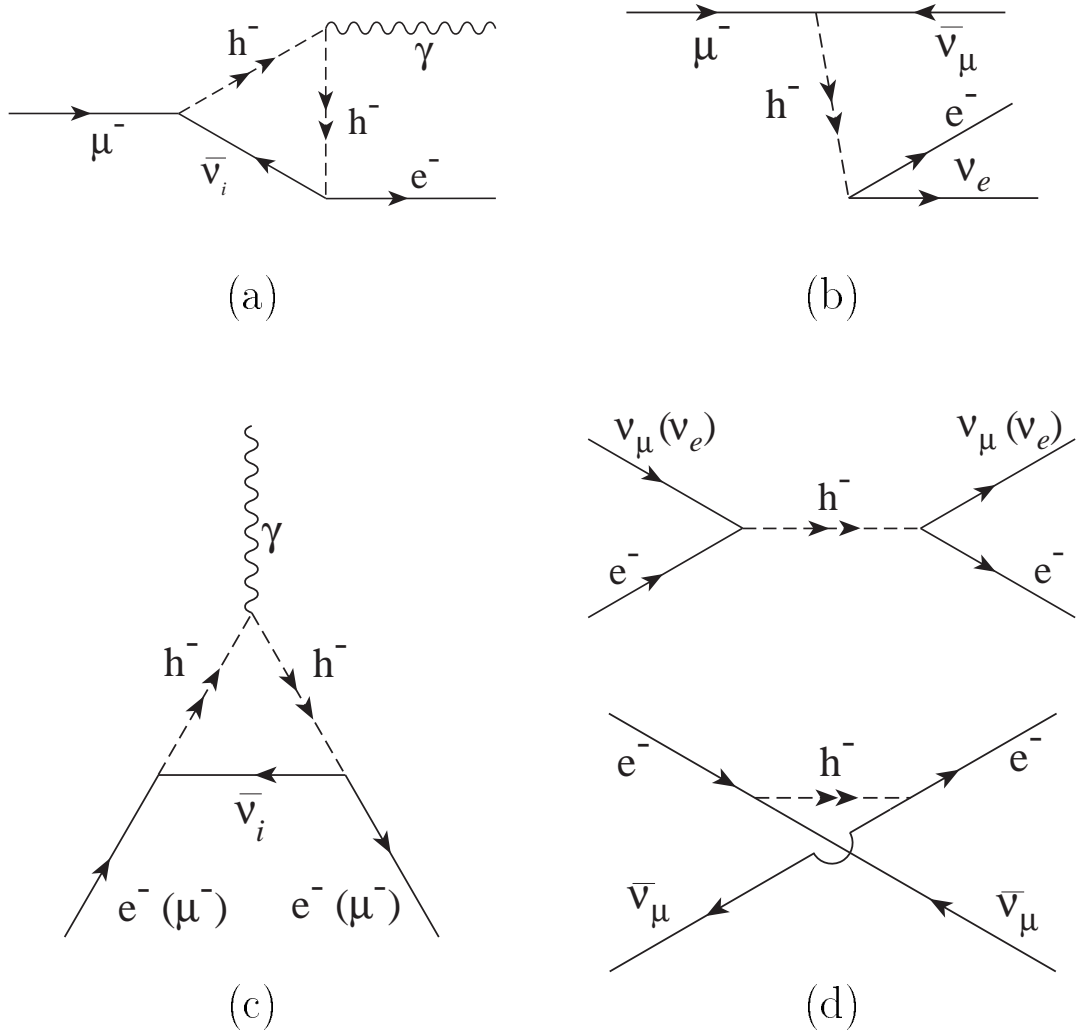
\item[\bf Fig. 2:] Diagrams contributing to the different
   observables described in the text:
   a) $\mu^-\rta e^- \gamma$;
   b) $\mu^-\rta\bb\nu_\mu e^- \nu_e$;
   c) anomalous magnetic moment of the electron and muon; and
   d) $\nu_\mu e$, $\bb\nu_\mu e$ and $\nu_e e$ collisions.
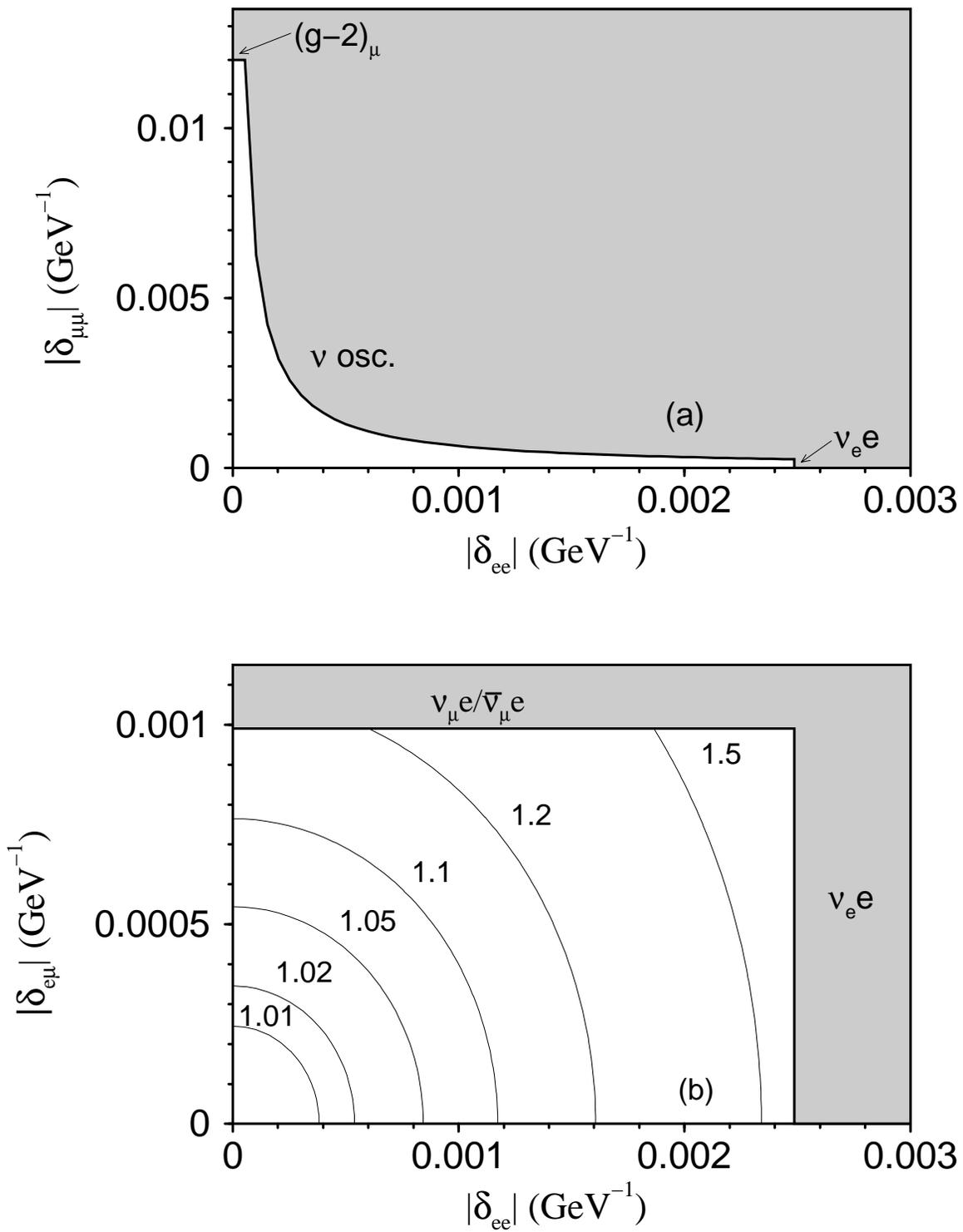
\item[\bf Fig. 3:] Allowed regions (white zones) in the
   $|\de{e e}|$-$|\de{\mu\mu}|$ plane (a) and $|\de{e e}|$-$|\de{e\mu}|$
   plane (b). In the latter we also show the contour lines for the ratio
   $R_{LCD}/R_{LCD}^{SM}$ as predicted by the Higgs triplet model.
\end{itemize}

\newpage
\begin{table}
\begin{center}
\begin{tabular}{|c|c|}
\hline
\multicolumn{1}{|c|}{90\%\ C.L. bound} &
\multicolumn{1}{c|}{Process}\\
\hline
\hline
$|\delta_{ee}| < 2.5\times 10^{-3}\, GeV^{-1}$   & $\nu_e e $ \\
$|\delta_{e\mu}| < 9.9\times 10^{-4}\, GeV^{-1}$ & $\nu_{\mu}e
                  \, , \, \bb\nu_{\mu}e $\\
$|\delta_{\mu\mu}| < 1.2\times 10^{-2}\, GeV^{-1}$ & $(g-2)_\mu $\\
$|\delta_{ee} \delta_{\mu\mu}| < 6.5\times 10^{-7}\, GeV^{-2}$ &
                  $\nu\,\mbox{osc.}$\\
\hline
\end{tabular}
\end{center}
\caption{Best bounds on $\delta_{ij}$.}
\label{tab1}
\end{table}

\vspace{3cm}

\begin{table}
\begin{center}
\small
\begin{tabular}{|c|c||c|c|}
\hline
\multicolumn{1}{|c|}{90\%\ C.L. bound} & \multicolumn{1}{c||}{Process}
& \multicolumn{1}{c|}{90\%\ C.L. bound} & \multicolumn{1}{c|}{Process}
\\
\multicolumn{1}{|c|}{(general)} & \multicolumn{1}{c||}{} &
\multicolumn{1}{c|}{(assuming $m_{h^+}<m_{h^{++}}$)} &
\multicolumn{1}{c|}{} \\
\hline
\hline
$|\delta'_{ee}| < 2.8\times 10^{-3}\, GeV^{-1}$   & Bhabha &
 $|\delta'_{ee}| < 2.5\times 10^{-3}\, GeV^{-1}$   & $\nu_e e $ \\
$|\delta'_{e\mu}| < 4.3\times 10^{-3}\, GeV^{-1}$ & $(g-2)_\mu $ &
 $|\delta'_{e\mu}| < 9.9\times 10^{-4}\, GeV^{-1}$ & $\nu_{\mu}e
                  \, , \, \bb\nu_{\mu}e $\\
$|\delta'_{\mu\mu}| < 4.3\times 10^{-3}\, GeV^{-1}$ & $(g-2)_\mu $ &
 $|\delta'_{\mu\mu}| < 4.1\times 10^{-3}\, GeV^{-1}$ & $(g-2)_\mu $\\
$|\delta'_{ee} \delta'_{\mu\mu}| < 8.6\times 10^{-6}\, GeV^{-2}$ &
                  $M\bb M$ &
 $|\delta'_{ee} \delta'_{\mu\mu}| < 6.5\times 10^{-7}\, GeV^{-2}$ &
                  $\nu\,\mbox{osc.}$\\
$|\delta'_{ee} \delta'_{e\mu}| < 2.3\times 10^{-11}\, GeV^{-2}$ &
                  $\mu\rta 3e$ &
 $|\delta'_{ee} \delta'_{e\mu}| < 2.3\times 10^{-11}\, GeV^{-2}$ &
                  $\mu\rta 3e$ \\
\hline
\end{tabular}
\end{center}
\caption{Best bounds on $\delta'_{ij}$.}
\label{tab2}
\end{table}

\end{document}